# Inclusive Child-centered AI: Employing design futuring for Inclusive design of inclusive AI by and with children in Finland and India

Sumita Sharma, Netta Iivari, Leena Ventä-Olkkonen, Heidi Hartikainen, and Marianne Kinnula

INTERACT Research Unit, University of Oulu, Finland. <firstname.lastname>@oulu.fi

Children increasingly use applications utilizing Artificial Intelligence / Machine Learning (AI/ML). Given the propensity of such applications to propagate existing social, gender, and racial biases, it becomes imperative to consider designing and developing child-centered AI applications for children. Furthermore, children should have opportunities and skills to critically reflect on current applications and envision and design better AI/ML applications that are ethical, specifically, those that are inclusive and fair. In our work, we focus on child-centered AI and inclusion. Using a two-fanged approach to inclusion and employing design futuring in our research with schools in India and Finland, children critically considered future technology design for all. In this paper, we present three cases of this work: a study with students at a school in New Delhi and two studies with students at schools in Oulu. Our work showcases how to design for inclusion - by designing for all, and how to design inclusively - by inviting children to envision the future, through design futuring approaches.

## 1 INTRODUCTION

Children increasingly use applications utilizing Artificial Intelligence / Machine Learning (AI/ML) in their everyday lives, from browsing social media to exploring content generation though Dall-e2 for AI Art and ChatGPT for text. However, these systems have a propensity to propagate existing social, gender, and racial biases (Keddell 2019; Van Schie and Adrianus 2022). One example of how this can impact children's futures is the UK A levels results in 2020, where an algorithm graded student performance based on three parameters: teacher's scores (predicted), past tests results, and school's historical performance. This resulted in grades that were several percentage points

______________________________________________________________________________________________

lower for students attending schools in low socio-economic regions, with many students missing the requisite grades for admission to their choice of university (Goodyear 2020). There are also studies on how recommendation systems used by social media apps such as YouTube, Instagram, and TikTok potentially promote mental health issues and challenges in adolescents (Barry et al. 2017). With such horrific examples already known, it becomes imperative to consider designing and developing AI applications utilizing child-centered approaches. UNICEF's Policy guidance on AI for children version 2.0 (UNICEF 2021), recommends nine-key requirements for child-centered AI, including "ensur(ing) inclusion of and for children" (#2), "Prioritiz(ing) fairness and non-discrimination for children" (#3), and "Prepar(ing) children for present and future developments in AI" (#8). Bearing these in mind, in our studies with schools in India and Finland, utilizing design futuring, children envision future technologies that are ethical – inclusive and fair.

Until now, design futuring has not been extensively explored with children in Child Computer Interaction (CCI research, while some seeds of such can be found (e.g., Iivari et al. 2022). Previous studies on design futuring with children, with its many flavors such as design fiction and speculative design (Dunne & Raby 2013), have focused on exploring sustainable environmental futures (Demneh & Darani 2020, Maxwell et al. 2019), newer forms of learning in future schools (Duggan et al. 2017), children's perceptions of and expectations from social robots (Rubegni et al. 2022), the ethics of emerging technologies (Hardy 2019, Kinnula et al. 2022), solutions to societal problems such as bullying (Ventä-Olkkonen et al. 2021) and on introducing Machine learning to children (Aki Tamashiro et al. 2021). The motivation for these studies is to make the world a better place for everyone by designing inclusive and fair (future) technologies (e.g., Kinnula et al. 2022, Sharma et al. 2022, Ventä-Olkkonen et al. 2021)

In the context of children and AI, however, current research is limited. Design futuring approaches entail envisioning (alternative) futures, questioning, imagining, and investigating different futures, preferably being emancipatory, critical, reflective in nature (e.g., Kozubaev et al. 2020). We see great potential in design futuring approaches in encouraging critical perspectives towards future technologies and technology practices and in providing opportunities to contribute to the design of future technologies.

In this study we scrutinize child-centered AI from the important viewpoint ethical AI and more specifically from the viewpoint of inclusion, addressing both inclusive AI solutions and inclusive design of them, utilizing design futuring approaches. We showcase the benefits of this two-pronged approach to inclusion – designing for inclusion and designing inclusively with children, with three empirical studies carried out in India and Finland.

## 2   DESIGNING FOR INCLUSION AND DESIGNING INCLUSIVELY

In our research, we worked with children in India and Finland, where children were invited to envision future technologies in their everyday lives: future classrooms (Case 1), solutions to reduce bullying at schools (Case 2), and future social robots (Case 3) (Table 1). In each of these, we employed design futuring approaches asking children questions such as "*what can or should your future robot do and not do?*" (Case 3), following up with reflections on the ethical implications of what they can and cannot do. Children also designed and prototyped their future imaginations using arts and crafts supplies (Cases 1, 2) and in the city center digital fabrication lab (Case 2).

Table 1. Overview of the three cases

| Participants | Method | Location | Main findings on inclusion and design futuring |
|---|---|---|---|
| **Case #1:** Imagining future teachers and friends in the classroom | | | |
| 15 children (ages 9-12; 6 girls, 9 boys) | 2-day hands-on workshops with arts and crafts activities, using Dall-e2 and Teachable Machines, reflections, discussions | In-person at a Fab Lab in Oulu, Finland | Inclusive design outcomes: several groups imagined future coins and cards available to all children at schools for financial independence, while other groups imagined freedom from everyday chores. *Inclusive design process*: children imagined future schools, envisioning how technologies feature in their future lives |
| **Case # 2:** Imagining solutions for solving the real-world problem of bullying | | | |
| 22 children (aged 11-12; 8 girls, 11 boys) and their teacher | 9 weekly workshops with hands-on activities, design fiction, arts and crafts, critical reflection, discussions, drama | In-person and online, at a school in Oulu, Finland with Fab Lab visits | *Inclusive design outcomes:* aiming at including those bullying and those bullied, building a happy school for all. *Inclusive design process:* an entire school class from a public school in Finland engaged in design futuring, inviting also their broader school community to reflect on the problem of bullying as part of the design process. |
| **Case #3:** Imagining future robots in everyday lives | | | |
| 31 children (aged 10-12; 14 girls, 17 boys) | A 3-hour workshop using speculative design activity, discussions, block programming the Nao v6 robot | In-person at a public school in New Delhi, India | *Inclusive design outcomes:* Do no harm' and 'do good' as intuitive ethical issues for children regarding robots. *Inclusive design process:* an entire school class from a public school in India engaged in design futuring. Hands-on experiences with robot shaped children's answers but provided grounding and inspiration to them. 'Strong anthropomorphizing of robots limiting children's thoughts. |

## 3 DISCUSSION AND CONCLUSION

With technology being so intertwined in children's everyday lives, children should have opportunities to envision and design future AI/ML applications that are ethical, that is, they are inclusive, fair, transparent, and trustworthy, as well as to have both opportunities and capabilities to critically reflect on the current technologies. We link this to the recent discussions on children's design capital (Mahboob Kanafi et al. 2021, 2022) and their role as 'design protagonists', i.e., children as active agents in their technology rich lives, reflecting on their own technology use as well as its use in the society (Iversen et al. 2017, Iivari & Kinnula 2018). We think design futuring approaches have a lot of potential for these purposes, especially such that embed a critical, reflective, and emancipatory stance (see Kozubaev et al. 2020). Ongoing research in CCI has not engaged much with design futuring approaches, however. We argue that such approaches can be used for engaging children in ethical AI design. In CCI, there is clear interest towards this type of approaches, several studies focusing on diversity, inclusion, equity, and empowerment (Iivari et al. 2022) and fostering critical transformation of the world (e.g., Antle et al. 2022, Iversen et al. 2017, Morales-Navarro et al. 2022, Ventä-Olkkonen et al. 2021, 2022). Children have also been engaged in critical analysis of digital technology from the perspective of its impacts on society (e.g., Antle et al. 2022, Iversen et al. 2017, Morales-Navarro et al. 2022, Schaper et al. 2020, Rubegni et al. 2022). We direct these critical approaches and stances towards children-centered AI and inclusion. Through our work, we highlight how to *design for inclusion* by designing for all and to *design inclusively* by inviting children to envision the future. We will further develop our two-fanged approach to inclusion, utilizing design futuring in hands-on workshops with children in schools in India and Finland, engaging children in critically considering future technology design for all.

## ACKNOWLEDGMENTS

This work is funded by the Academy of Finland (Grant #324685 Make-a-Difference project, Grant #340603 PAIZ project, Grant #351584 Making sense of 'technology protagonism' project) at the University of Oulu. It is connected to the GenZ project, a strategic profiling project in human sciences at the University of Oulu, supported by the Academy of Finland (Grant #318930, Profi4) and the University of Oulu.